\documentclass[a4paper,11pt]{article}
\pdfoutput=1 
\usepackage{xcolor}
\usepackage{comment}
\newcommand{\be}{\begin{eqnarray}}
\newcommand{\ee}{\end{eqnarray}}

\usepackage{soul}
\usepackage{cite}
\usepackage{jheppub} 

\usepackage[T1]{fontenc} 

\title{The Gross-Pitaevskii equations of a static and spherically symmetric condensate of gravitons}

\author[a]{Francesc Cunillera}
\author[b]{and Cristiano Germani}

\affiliation[a]{Departament de F\'isica Qu\`antica i Astrof\'isica, Facultat de F\'isica, Universitat de Barcelona, Mart\'i i Franqu\`es 1, 08028 Barcelona, Spain}

\affiliation[b]{Institut de Ci\`encies del Cosmos, Universitat de Barcelona, Mart\'i i Franqu\`es 1, 08028 Barcelona, Spain}

\emailAdd{frcunilg8@alumnes.ub.edu}
\emailAdd{germani@icc.ub.edu}

\abstract{{In this paper we consider the Dvali and G\'omez assumption that the end state
of a gravitational collapse is a Bose-Einstein condensate of gravitons. We then construct the two Gross-Pitaevskii equations for a static and spherically symmetric configuration of the condensate. These two equations correspond to the constrained minimisation of the gravitational Hamiltonian with respect to the redshift and the Newtonian potential, per given number of gravitons. We find that the effective geometry of the condensate is the one of a gravastar (a DeSitter star) with a sub-Planckian cosmological constant, for masses larger than the Planck scale. Thus, a condensate corresponding to a semiclassical black hole, is always quantum and weakly coupled. Finally, we obtain that the boundary of our gravastar, although it is not the location of a horizon, corresponds to the Schwarzschild radius.}}

\begin{document} 
\maketitle
\flushbottom
\section{Introduction}

Assuming that black holes can be described by a Bose-Einstein condensates (BEC) of gravitons, in a seminal paper, Dvali and G\'omez \cite{dvali} have shown that many of the quantum properties of semiclassical black holes might be simply connected to the number of gravitons $N$ forming the condensate. In particular, in this picture, black holes are supposedly constituted by weakly coupled ``gravitons'' with an averaged wavelength of the black hole size. {Henceforth, we identify ``graviton'' as the metric of the effective geometry we will find for the condensate, which is subject to Planck's quantisation hypothesis} (see the later eq. \ref{planck_quant}).

Most of the activity on this idea has been so far focused on reproducing the known semiclassical results of a large black hole outside its own Schwarzschild radius. For example, in \cite{dvali} the Bekenstein-Hawking entropy has been shown to be uniquely determined by $N$. In \cite{hawking,hawking2}, the same Authors have related the Hawking evaporation to the assumption that the BEC of gravitons sits on a quantum critical point. In this case, the Hawking evaporation is nothing else than the depletion of the condensate that, for large ``black holes'', adiabatically change its own size. Then, when a BEC ``black hole'' loses half of its mass, at the Page time \cite{page}, the adiabaticity of this process is broken and $1/N$ corrections, rather than $e^{-N}$ as assumed in any typical semiclassical approximation, may become important and unitarity can be in principle restored \cite{entropy}. 

Recently, those works were complemented in \cite{domenec} with the first tentative study of the BEC of gravitons, the would be black hole interior. There, the quantum fluctuations on a Schwarzschild background were assumed to condensate. In this case, inspired by quantum mechanics, the particle number constraint was implemented into the action as a ``mass'' term for the fluctuations, where the mass was identified as the chemical potential of the condensate. 

Although we were motivated by this pioneer analysis, we have, however, taken a completely different direction. First of all we have not assumed the existence of any background solution on which a condensate of gravitons sits. In other words, in this work we put forward the original idea that the background itself is quantum, i.e. a condensate of gravitons.
Secondly, noticing that a BEC of gravitons spontaneously breaks diffeomorphism invariance, we have implemented the particle number constraint not in a ``covariant form'', as done in \cite{domenec}, but rather in a form that is invariant only under the unbroken symmetries: time translation and rotational invariance. In addition, in our approach, we avoided possible drawbacks related to the introduction of a ``mass'' term for the graviton (see e.g. \cite{nico}). 

Generically, 
the constrained variational equation
	\begin{equation}
	\delta H -\mu\delta N=0\ ,
	\label{variation}
	\end{equation} 
where $H$ is the Hamiltonian, $N$ the number of fields and $\mu$ the chemical potential, {leads to the Gross-Pitaevskii (GP) equation under the assumption that all particles are in the same quantum state. From a different perspective,} \eqref{variation} {is equivalent to the minimisation of a grand potential when considering that all particles are identical}\footnote{{There have been some approaches to treating black holes as a grand canonical ensemble }\cite{ge1,ge2}{ from the point of view of relating the thermodynamical black hole properties to the statistical ensemble in GR.}}. 

A striking difference between the non-relativistic GP case and gravity is that, in the former, the dynamics is fully obtained by minimising the Hamiltonian in terms of a single (wave) function while in the latter the Hamiltonian is described by two, rather than one, potentials. In fact, in General Relativity (GR), the gravitational system produced by a massive body is determined by two quantities: the Newtonian potential (related to space distortions) and the redshift (related to time distortions). In the weak limit of classical GR the two coincide, however, generically, they are independent quantities. Thus, a BEC of a static spherically symmetric ball of gravitons is described by two {generalised} GP equations {(that we will refer to as GP equations for short)} obtained by minimising \eqref{variation} with respect to the Newtonian potential and the redshift.

\section{Constructing and solving the GP equations}

Consider a static spherically symmetric configuration. Keeping explicit rotational invariance, the metric can be gauged as  
	\begin{equation}
	{ds^2=-L(r)^2dt^2+\frac{dr^2}{1-2 \frac{G M(r)}{r}}+r^2d\Omega^2\ ,}
	\label{metric}\end{equation}
where $L$ is the lapse function parameterising time translations (related to the redshift), $M(r)$ parameterise the curvature of the spatial volume (related to the Newtonian potential) {and $G$ is Newton's constant}. 

In this case, by using the ADM formalism, we can write down the gravitational Hamiltonian of a static spherically symmetric ball as \cite{ADM} 
	\begin{equation}
	\begin{split}
	{H_G=-\frac{1}{4 G}\int dr\ r^2{1\over\sqrt{1-2 \frac{G M(r)}{r}}}L\ {\cal R}\ ,}
	\end{split}
	\label{ham_grav}
	\end{equation}
where $\cal R$ is the three-dimensional Ricci scalar constructed from the three-dimensional metric. {Finally, in} \eqref{ham_grav} {we have dropped eventual boundary terms. We shall further discuss about these terms at a later time.}

With the metric \eqref{metric} we then have
\be
{H_G=-\int dr {M'\over\sqrt{1-2 \frac{G M(r)}{r}}}L \ ,\label{hamiltonian}}
\ee
where $M'\equiv \partial_r M$.

 In order to do the constrained variation \eqref{variation}, we need to define the number of particles $N$ and the chemical potential $\mu$. 
	
Following \cite{dvali}, we assume that the BEC is formed by $N$ gravitons that have in average the same wavelength $\lambda_g$. This also corresponds to the physical radius of the condensate so that its volume is fixed to be $V=4\pi \lambda_g^3/3$. Note that, because of the spatial distortion, fixing this volume does not yet fix the coordinate radius of the condensate $r_c$. We will parameterise it as $r_c\equiv \alpha r_s$, where $\alpha$ is a constant to be determined and $r_s$ is the would-be Schwarzschild radius. As in \cite{dvali}, we expect $\lambda_g$ to be also parameterised by $r_s$. Therefore, we define $\lambda_g=\kappa r_s$.

We then finally have that the number of particles of the condensate is 
\be
N=\langle E\rangle \lambda_g\ ,
\label{planck_quant}
\ee 
where $\langle\cdot\rangle$ stands here for the spatial average and $E$ is the energy of the condensate. Because of the redshift, the condensate energy is $E=L E_{\infty}$ where $E_{\infty}$ is the energy as seen from infinity. To easily compare our findings to the BH case, we will fix $E_\infty\equiv M_{\infty}\equiv \dfrac{r_s}{2G}$, where $M_{\infty}$ is the would be black hole mass related to its own radius.

Thus, we have
\be
{N=\frac{3}{8\pi G \kappa^2 r_s}\int dV L=\frac{3}{2 G \kappa^2 r_s}\int dr\ r^2 {L\over\sqrt{1-2 \frac{G M(r)}{r}}}\ .}\label{N}
\ee
Note that the same result can be found from a slightly different perspective. Because the condensate is at rest, it has a total momentum $P^\mu=\{M_{c},\vec 0\}$. Assuming a uniform radial distribution of mass, the momentum density ${\cal P}^{\mu}$ is easily found to be ${\cal P}^\mu=\frac{P^\mu}{V}$. The energy density of the condensate, containing a redshift factor, is ${\cal E}=\sqrt{-{\cal P}^\mu{\cal P}_\mu}$ and so the total energy $E=\int {\cal E} dV$. Given that the BEC of gravitons is formed by $N$ iso-energetic particles of energy $\lambda_g^{-1}$, we have that $N=E\lambda_g$, which is \eqref{N}.

We are now only left to define the chemical potential. As usual, in GP equations the chemical potential is determined fully by the boundary conditions imposed. This is indeed what we will find. However, we can already obtain a rough order of magnitude of it following the discussion of \cite{dvali}. 

The chemical potential is the energy one needs to extract a graviton from a BEC. This was estimated to be $\mu\sim \frac{1}{\lambda_g}$ in \cite{dvali}. Therefore, we define
	\begin{equation}
	{\mu\equiv-\frac{\gamma}{\lambda_g}}
	\label{mu}
	\end{equation} 
where $\gamma$ is a numerical factor that cannot be found from the analysis of \cite{dvali}. Note that, although $\gamma$ could have any sign, we have already put a minus sign in \eqref{mu} since we expect the chemical potential to be negative, as in any typical BEC. We will indeed see that this is the case.

To simplify the notation we define the constant ${\beta^2\equiv \frac{3 \gamma}{2 G \kappa^3 r_s^2}}$ so that we finally obtain
\be
{\mu N=-\beta^2\int dr\ r^2 {L\over\sqrt{1-2 \frac{G M(r)}{r}}}\ .}
\ee
Suppose now $\mu N=0$. Then the variation \eqref{variation} would lead to a  Schwarzschild geometry with radius $r_s$ as a unique solution. By looking at $\beta$, we would then be tempted to state that a very large BEC, with $r_s\rightarrow\infty$, would be extremely classical. However, that is a premature conclusion. Supposing that the graviton wavelength is of order $r_s$, we can already expect that the total number of graviton is proportional to the volume of the condensate itself ($\sim r_s^3$). Thus, if our expectations are correct, $\mu N$ would actually grow with the volume\footnote{{In the standard BEC case, by increasing the volume, and keeping the number of particles fixed, one would dilute the BEC and evolve into a classical gas. In our case however, because of the maximal packing, the volume will be tied to $N$ and this change of state will never happen.}}. In the following we will indeed prove these expectations.

\subsection{Solving the GP equations}
Let us start by considering the variation \eqref{variation} with respect to the lapse $L$. We obtain the first GP equation
\be
{M'=\beta^2 r^2\ .}\label{eqL}
\ee
{In principle, to obtain} \eqref{eqL} {one has to drop a boundary term, which is equivalent to adding a Gibbons-Hawking (GH) term to} \eqref{hamiltonian}{, as similarly discussed in }\cite{Hawking_hamiltonian}. {This, as we shall see, defines the boundary energy of the condensate and eventually fixes all parameters of our system.}

By inspecting the equation \eqref{eqL} we see that the condensate effectively behaves as a star of constant density ${\rho={\beta^2}}$. The solution of the above equation is
\be
{M(r)=\frac{\beta^2}{3} r^3=\frac{1}{2G}\frac{\gamma}{\kappa^3 r_s^2}r^3\ ,}\label{Ms}
\ee
where we have fixed the constant of integration such that the geometry is regular at the centre of the ball. This choice complies with the initial assumption that the condensate is a bound state of weakly coupled gravitons. 

One might wonder whether the same choice could have be done in GR by considering a ball with a Minkowski, instead of Schwarzschild, geometry. In that case the matching to a curved exterior, making the overall solution non-trivial, would have been not possible. On the contrary, in our case we will be able to match our interior (the regular ball) with a non-trivial exterior geometry.  

Making now the variation with respect to the mass $M(r)$ of \eqref{variation}, the second GP equation will be
\be
{\frac{L'}{g_{rr}} -\frac{G M}{r^2}L=-{\beta^2} {G r} L\ .} 
\ee
Substituting $M$ from \eqref{Ms} we then finally find
\be
{\frac{L'}{g_{rr}}+\frac{2}{3}\beta^2  G r L=0\ .}\label{M}
\ee
The solution of \eqref{M} is, as in the Schwarzschild case,
\be
L^2=\frac{1}{g_{rr}}\ ,
\ee
where the constant of integration has been re-absorbed into a time reparameterization.

Carry on our comparison with a constant density star, we see that this solution corresponds to a star with also a constant (negative) pressure $p=-\rho$. In other words, the BEC effective metric is the one of DeSitter in static coordinates, with cosmological constant\footnote{The emergence of an effective cosmological constant in particular (gravity unrelated) BECs, has been also noticed in \cite{florian}.} 
\be
{\Lambda_{eff}=2G\beta^2=\frac{3\gamma}{\kappa^3r_s^2} \ .} \label{Lambda}
\ee
This kind of star has been dubbed {\it gravastar}\footnote{We thank Jorge Rocha for noticing this.} in \cite{mottola} and for this reason we will from now on call the BEC of gravitons a gravastar. {Note that the emergence of the ``cosmological constant'' might only be a coincidence of the static case. The reason is that, in principle, the way we constructed $N$ is not diffeomorphism invariant, as explained before.}

The geometry we have found so far agrees with our earlier expectations that the product $\mu N$ grows with the BEC volume: for a finite graviton wavelength, there is no parameter tuning we can make such that our gravastar becomes a black hole. Therefore, our gravastar can {\it never} be a black hole.

We conclude this section by noticing that the emergence of a cosmological constant could also be expected. Generically, a BEC condensate of massless particles generates an effective mass for the fluctuations\footnote{We thank Tomeu Fiol for pointing this out.}. In this sense, the assumption of \cite{domenec} that the graviton fluctuations on a ``Schwarzschild condensate'' have a mass, well resonate to this fact. In our case, however, the expected non-derivative gravitons interaction, loosely speaking a ``mass'' term\footnote{Note that although a cosmological constant is a non-derivative interaction for the gravitons, at least in GR, it is strictly speaking not a mass term as it does {\it not} break diffeomorphism invariance.}, is generated by the emergent cosmological constant rather than being imposed as in \cite{domenec}.

In the next section we will fix the parameters of the gravastar by giving the boundary conditions of our GP problem.

\subsection{Matching with the exterior and boundary energy}\label{boundary_energy}

Because we are considering an asymptotically flat space with Coulomb flux (Newtonian potential) as the gravastar exterior, the wavelength of the average graviton outside the gravastar is infinite and so the product $\mu N=0$ there. Therefore, the exterior solution matching the gravastar effective metric is Schwarzschild, i.e.
\be
ds_e^2=-(1-\frac{r_s}{r})dt^2+\frac{dr^2}{1-\frac{r_s}{r}}+r^2d\Omega^2\ ,
\ee 
where ``$e$'' stands for exterior.

The two solutions, the exterior and the interior, are matched in the coordinate radius $r_c=\alpha\ r_s$. We then find,
\be
{\left.g_{rr}^{\ i}\right|_{r_c}=\left.g_{rr}^{\ e}\right|_{r_c} \quad \rightarrow  \quad \left.\frac{\gamma}{\kappa^3}\frac{r^2}{r_s^2}\right|_{r_c}=\left.\frac{r_s}{r}\right|_{r_c}\quad \rightarrow \quad \kappa^3=\alpha^3 \gamma\ ,}
\ee
where ``$i$'' stands for interior. This implies
\be
ds^2_i=-\left(1-\frac{1}{\alpha^3}\frac{r^2}{r_s^2}\right)dt^2+\frac{dr^2}{1-\frac{1}{\alpha^3}\frac{r^2}{r_s^2}}+r^2d\Omega^2\ ,
\ee
The extra condition we have to impose is that the spatial volume of the BEC of the gravitons is $\frac{4\pi}{3}\lambda_g^3$. This leads to the transcendental equation
\be
\gamma=\frac{3}{2}\alpha^{3/2}\left(\arcsin\left(\frac{1}{\sqrt{\alpha}}\right)-\frac{\sqrt{\alpha-1}}{\alpha}\right)\ .\label{g}
\ee
Then, both $\gamma$ and $\kappa$, parameterising respectively the scattering strength between gravitons in the condensate and the physical wavelength of the averaged graviton, are determined by $\alpha$.

We immediately see from \eqref{g} that, because $\gamma$ is real, we get
\be
\alpha\geq 1\ .
\ee 
Note that even when $\alpha=1$, i.e. the would be Schwarzschild horizon coincides with the condensate boundary, the location $r=r_s$ is not an event horizon, although it is an infinite redshift surface. This is due to the fact that there is no change of causality structure by crossing the gravastar boundary. 

 {In addition to the pressure energy at the boundary, one also has a contribution due to the Gibbons-Hawking term\footnote{{We thank the anonymous referee for noticing this.}}. Following} \cite{Hawking_hamiltonian}{, one can write the energy due to the GH term at the boundary surface $S$, for a compact geometry, as}\footnote{{Note that in }\cite{Hawking_hamiltonian}{, the lapse function, as is usual in the ADM formalism, is denominated $N$. However, we have renamed it $L$ to avoid confusions between the particle number and the lapse function.}}
\be
{E_{GH}=\frac{1}{8\pi G} \oint_S {}^2K\ ,}
\ee
{where ${}^2K$ is the trace of the two dimensional extrinsic curvature ${}^2K_{\mu\nu}$ on the boundary $S$. In particular for our case, $S$ are a family of surfaces at constant radius that foliate the 3-spatial manifold. Thus, ${}^2K_{\mu\nu}$ runs over the two angular components and takes the form}
\be
{{}^2K_{\mu\nu}=\frac{1}{2\tilde{L}}q_{\mu\nu,r}\ ,}
\ee
{with $q_{\mu\nu}$ the induced metric on the two dimensional boundary surfaces $S$ and $\tilde{L}$ is the associated ``lapse function'' to the 1+2 splitting of the spatial metric.} 

{Then, the energy at the boundary is}
\be
{E_{GH}=\left.\frac{L}{G} \ r\right|_{r_c}=2 M_\infty \alpha \sqrt{1-\frac{1}{\alpha}} \ .}
\ee
{Requiring this energy to be compatible with the energy of the condensate introduced previously evaluated at the boundary, i.e. $E=M_\infty \sqrt{1-\frac{1}{\alpha}}$, immediately implies that $\alpha=1$\footnote{One might argue that $\alpha=1/2$ is a consistent solution as well, however that would not allow the matching with the exterior and would set an imaginary scattering amplitude for the graviton.}. This is actually what one should expect. Indeed, the fact that the radius of the condensate is the same of the Schwarzschild radius, implies that the exterior is well described by GR, as we have also discussed earlier.}

{Finally then, our system is completely determined with}
\be
{\alpha=1\ ,\qquad \gamma=1.33\ ,\qquad\kappa=1.10\ .}
\ee
{It is interesting to remark that the expression for the energy vanishes exactly at the boundary. This is reminiscent of the Hawking temperature becoming infinite at the horizon of the BH, i.e. }${T_H\propto E^{-1}\Big |_{r \rightarrow r_s}\propto (M L)^{-1}\Big |_{r\rightarrow r_s}\rightarrow \infty\ .}$

\subsection{The effective energy-momentum tensor of the BEC}

We have found that the BEC condensate of spherically symmetric and static gravitons is nothing-else than a DeSitter universe with a cosmological constant given by 
\be
{\Lambda_{eff}=\frac{3}{r_s^2}\lesssim \frac{{\cal O}(1)}{L_p^2}\ \mbox{  for  }\ r_s\gtrsim L_p\ ,}
\ee 
{Where $L_p$ is the reduced Planck Length. Thus, for masses slightly larger that the Planck scale, the gravastar curvature is always sub-Planckian and so our treatment in terms of a ball of weakly coupled gravitons is fully justified. This is a striking difference between the BEC ball of graviton and a Schwarzschild black hole.}

In addition to the energy density of the BEC ball, the gravastar has also a surface energy due to the fact that the first derivatives of the metric ``jump'' between the interior and the exterior. In other words, the pressure of this star is not continuous across its surface and thus an observer falling into the ball will experience a ``wall''. This might be connected to the firewall of \cite{firewall} or might be related to the fact that the gravastar wants to release energy in terms of a Hawking-like evaporation, we leave the study of these issues for a future research.   

\section{Conclusions}
In this paper{, assuming its existence,} we have studied the physics of a static and spherically symmetric Bose-Einstein condensate of gravitons. We have found that the effective geometry of this condensate is the one of a weakly curved gravastar with a cosmological constant inversely proportional to the square of the total mass. Thus, a Bose-Einstein condensate of gravitons is never ``classical'', although its exterior geometry is the one of Schwarzschild. {The matching point of the gravastar to the Schwarzschild geometry is at a coordinate radius equal to the would be Schwarzschild radius.} However, the gravastar has no horizon even if it has an infinite redshift surface. {This well complies with the intuition that the exterior of a large ``black hole'' is well described by semiclassical gravity.}

{A question of particular interest is whether or not the DeSitter solution we found is stable. In principle, one expects that a model of quantum BH would contain an instability mimicking Hawking's evaporation. Thus, if a BH can be modelled as a BEC, the condensate we have found must be unstable, as also discussed in} \cite{Flassig_nico}{. The answer to this question deserves further studies that we leave for future research.}

\acknowledgments
F.C. and C.G. would like to thank Gia Dvali, Dom\`enec Espriu, Tomeu Fiol, David Pere\~n\'iguez and Nico Wintergerst for discussions and Vitor Cardoso and Paolo Pani for correspondence. C.G. is supported by the Ram\'on y Cajal program and partially supported by the Unidad de Excelencia Mar\'ia de Maeztu Grant No. MDM-2014-0369 and by the national grants: FPA2013-46570-C2-2-P and FPA2016-76005-C2-2-P.

\end{document}